\documentclass[12pt,journal,onecolumn,draftcls]{IEEEtran}
\usepackage{epsfig}
\usepackage{times}
\usepackage{float}
\usepackage{afterpage}
\usepackage{amsmath}
\usepackage{amstext}
\usepackage{amssymb,bm}
\usepackage{latexsym}
\usepackage{color}
\usepackage{graphicx}
\usepackage{amsmath}
\usepackage{amsthm}
\usepackage{graphicx}
\usepackage[center]{caption}
\usepackage{pstricks}
\usepackage{subfigure}
\usepackage{booktabs}
\usepackage{multicol}
\usepackage{lipsum}
\usepackage{dblfloatfix}
\usepackage{url}
\usepackage{epstopdf}
\allowdisplaybreaks

\newtheorem{thm}{Theorem}

\allowdisplaybreaks

\newcommand{\dof}{\mathsf{DoF}}

\DeclareMathOperator{\rk}{rk}
\DeclareMathOperator{\nul}{Null}

\begin{document}

\title{On the DoF region of the two-user Interference Channel with an Instantaneous Relay}

\author{
\IEEEauthorblockN{Tang Liu$^*$, Daniela Tuninetti$^*$, and Sae-Young Chung$^{\dagger}$}\\
$^*$ University of Illinois at Chicago, Chicago, IL 60607 USA,\\
Email: {\tt tliu44, danielat@uic.edu}\\
$^{\dagger}$ KAIST, Daejeon, 305-701, South Korea,\\
Email: {\tt sychung@ee.kaist.ac.kr}
\thanks{This work was presented in part at ISIT 2015.
The work of T.~Liu and D.~Tuninetti was partially funded by NSF under award number 1218635; the contents of this article are solely the responsibility of the author and do not necessarily represent the official views of the NSF.
The work of S.-Y.~Chung was supported in part by the CISS through the Global Frontier Project.}
}\maketitle

\begin{abstract}
This paper studies the Degrees of Freedom (DoF) of the two-user multi-antenna Gaussian interference channel with an {\em instantaneous relay}, or relay without delay, where the relay transmitted signal in channel use $t$ can depend on all received signals up to and including that at channel use $t$. 
It is assumed that the two transmitters and the two receivers have $M$ antennas, while the relay receives through $N$ antennas and transmits through $L$ antennas. 
An achievable DoF region is derived, for all possible values of $(M,N,L)$, based on a memoryless linear transmission strategy at the relay that aims to {\it neutralize} as much interference as possible at the receivers.
The proposed scheme is shown to attain the largest sum DoF among all memoryless linear transmission strategies at the relay and to actually be optimal for certain values of $(M,N,L)$.
\end{abstract}


\section{Introduction} 
\label{sec:introduction}
The Interference Channel (IC) has attracted a lot of attention recently due its practical relevance in wireless networks.
The central question is how to deal cleverly with the interference instead of simply avoiding it or treating it as noise, as currently done in commercial networks. 
The breakthrough idea of Interference Alignment (IA) was presented in~\cite{ic-ia} as a way to ``consolidate'' interference in a lower dimensional space of the received space at each receiver. 
IA was shown to achieve the optimal sum degrees of freedom (DoF) for the IC for any number of active users and for the first time showed that Gaussian IC are not intrinsically interference limited~\cite{ic-ia}.

IA has found many applications beyond the classical IC setting.
For example, relays can play an important role in IA-based scheme.
In~\cite{IC-IA-relay}, it was shown that joint beam-forming between the relay and the transmitters makes IA feasible without channel state information at the transmitters.
In multi-hop scenarios~\cite{2by2by2}, the relays can cooperate to perform interference neutralization at the receivers. 
In spite of these improvements brought by relays, it is known that conventional (strictly causal) relaying can not provide DoF gain for fully connected ICs with generic channel matrices~\cite{relay-not-increase-dof}. However, this is not true for a novel type of relay called {\it Instantaneous Relay} (IR). 

IRs were first introduced in~\cite{relay-wo-delay} as a model where the signal transmitted by the relay in channel use $t$ can depend on all received signals up to and including that at channel use $t$. This is in contrast to conventional relaying, where the transmitted signal can depend on previously received signals in a strictly causal fashion.
IRs have been shown recently to be a special case of channels with in-block memory~\cite{iBM}, which generalize classical memoryless networks.

In channel with a single transmitter-receiver pair, the rate achieved with an IR is in general larger than the one achieved with a conventional relay~\cite{relay-with-delays}. 
For the two-user IC with an IR, outer bounds were derived in~\cite{strong-verystrong} and shown to be achievable by amplify-forward relaying in a non-asymptotic way for strong and very strong interference scenarios.

In~\cite{instantaneous-relaying}, the uninformed non-cooperative (where the transmitter-receiver pairs are not aware of the existence of an IR) and the informed cooperative $K$-user IC with an IR were studied.
It was shown that the IR improves the achievable rate region and provides better user-fairness in both scenarios  compared to the classical IC setting.

In~\cite{causal-relay-networks} general networks with IRs were investigated. Depending on whether the IRs have their own messages or not, two different cut-set bounds were proposed. The bounds were proved to be tight for the causal vector Gaussian two-way relay channel and the causal vector Gaussian relay channel.

Besides rate, IRs markedly differ from conventional relays as the can provide strict DoF gain compared to the case of absence of relays.
The interference aligned neutralization scheme proposed in~\cite{2ic-ir} achieves $3/2$ sum DoF for two-user IC with an IR and $5/3$ sum DoF for two-user $X$ channel, as opposed to $1$ sum DoF for two-user IC and $4/3$ sum DoF for two-user $X$ channel with conventional relays (in which case the sum DoF is as if the conventional rely is not present). 
The authors in \cite{2way_IC_IR} demonstrated a DoF optimal (i.e., achieving the cut-set upper bound) scheme for the two-way $K$-user IC with a multiple antenna IR having $2K$ antennas.


\subsection{Contribution} 
\label{ssub:contribution}
In this paper we study the two-user IC with one IR. In our model, the two transmitters and the two receivers have $M$ antennas, while the relay receives through $N$ antennas and transmits through $L$ antennas. The case $N=L=M$ was studied in~\cite{2ic-ir}, where it was shown that it is possible to achieve $3M/2$~sum DoF, which is strictly greater than $M$~sum DoF achievable with a conventional relay.
In this paper not only we study the achievable DoF region of the general $(M,N,L)$ case but, as opposed to~\cite{2ic-ir}, we also provide outer bounds.

Our main contribution is two-folded.
First we propose an achievable scheme for all possible values of $(M,N,L)$.
Then we show that our scheme is not only optimal among all possible linear transmission strategies at the IR, but it attains the optimal DoF region of the channel for $\max\left\{N,L\right\} \geq 2\min\left\{M,N,L\right\}$. 
Our achievable scheme aims to neutralize as much interference as possible at the receivers.
Our information theoretic outer bounds are derived through a genie argument so as to enhance the original channel to one whose DoF is known.
The optimality among  linear strategies is proved by generalizing the DoF of the classical $2$-user IC without relay when the channel matrices may be rank-deficient~\cite{rk-deficient-mimo}, which is a result of interest in itself.



\subsection{Paper Organization} 
\label{ssub:paper_organization}
Section \ref{sec:model_setup} introduces the channel model and states the main result of this work. 
The achievable DoF region is analyzed in Section \ref{sec:achievability}.
Two types of converse are discussed in Section \ref{sec:converse}.
Section \ref{sec:conclusion} concludes the paper.
Some proofs can be found in Appendix.

\subsection{Notation} 
\label{ssub:notation}
In the paper we use the notation convention of~\cite{NIT}.
For a matrix $M$, we use $\rk(M)$ to indicate its rank, 
$\text{span}\{M\}$ for its coloum span, 
$M^{\dagger}$ for its pseudo inverse, 
$\nul\left(M\right)$ for its null space, 
$\nul_{r:n}\left(M\right)$ for the first $n$ rows of $\nul\left(M\right)$  and
$\nul_{c:n}\left(M\right)$ for the first $n$ columns of $\nul\left(M\right)$. 

\section{Channel Model and Main Result} 
\label{sec:model_setup}

\begin{figure}
\centering
\includegraphics[width=0.5\columnwidth]{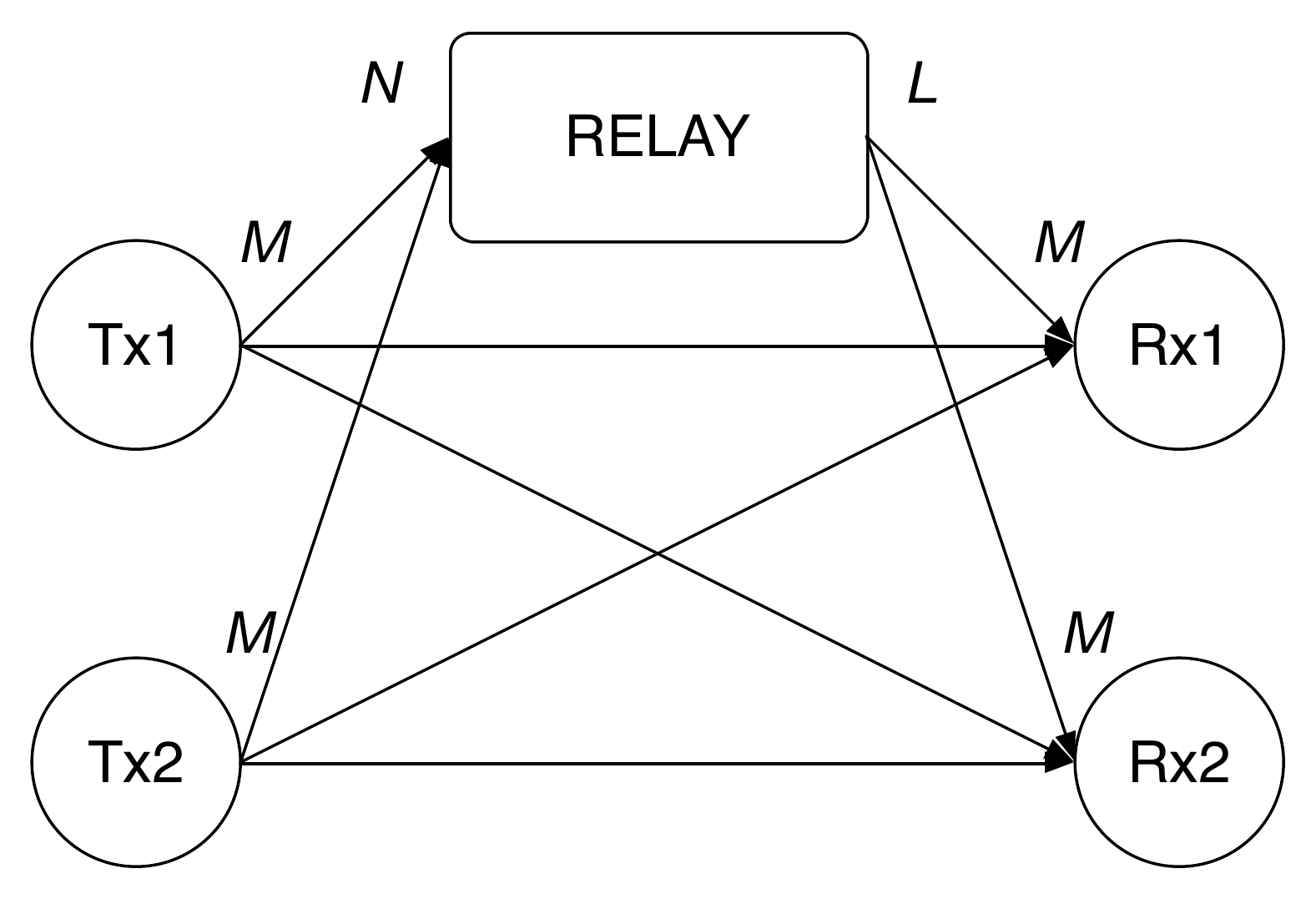}
\caption{Channel model.}
\label{fig:channelmodel}
\end{figure}

We consider the Gaussian IC with two transmitter-receiver pairs (nodes $1$ and $2$)
and one IR (node $0$), or 2IC+IR for short in the following.
The channel model is shown in Fig.\ref{fig:channelmodel}.
Each transmitter and receiver has $M$ antennas while the relay has $N$ antennas for receiving and $L$ antennas for transmitting.
The received signal in one time slot at the relay is
\begin{subequations}
\begin{align}
Y_{0} &=H_{01}X_{1}+H_{02}X_{2}+Z_{0} \in \mathbb{C}^{N \times 1}, 
\end{align}
and the received signals in one time slot at the receivers are
\begin{align}
Y_{1} & =  H_{11}X_{1}+H_{12}X_{2}+H_{10}X_{0}+Z_{1}\in \mathbb{C}^{M \times 1}, \\
Y_{2} & =  H_{21}X_{1}+H_{22}X_{2}+H_{20}X_{0}+Z_{2}\in \mathbb{C}^{M \times 1},   
\end{align}
where 
$X_{i}\in \mathbb{C}^{M \times 1}$ is the transmitted signal by transmitter $i\in[1:2]$ and
$X_{0}\in \mathbb{C}^{L \times 1}$ is the transmitted signal by the relay, 
subject to the power constraint $\mathbb{E}[\|X_i\|^2]\leq \mathsf{P}, i\in[0:2]$; 
the noises $Z_{i}, i\in[0:2],$ are independent circularly-symmetric complex Gaussian vectors with zero mean and covariance matrix equal to the identity;
each entry of the channel matrices $H_{ij}, (i,j)\in[0:2]^2, (i,j)\not=(0,0),$ are drawn in an i.i.d. fashion from a continuous distribution (i.e., are generic \cite{ic-ia}) and known to all nodes.
\label{eq:channeloutputs}
\end{subequations}

transmitter~$i$ has an independent message $W_i$, uniformly distributed on $[1:2^{nR_i}]$, for receiver~$i$ where $n$ is the block-length and $R_i$ is the rate in bits per channel use for user pair $i\in[1:2]$; the transmitted codewords are $X^n_i(W_i), \ i\in[1:2]$. The relay sends $X_{0,t}(Y^{t}_{0})$  at time $t\in[1:n]$. receiver $i$ outputs the estimate $\widehat{W}_i(Y_i^n), \ i\in[1:2]$. 
A rate pair $(R_1,R_2)$ is achievable if there exists a sequence of codes such that $\Pr[\widehat{W}_i(Y_i^n)\neq W_i], \ i\in[1:2],$ goes to zero as $n\to\infty$. The capacity region is the convex closure of all achievable rate pairs.

In this paper we are interested in characterizing the DoF of the channel, which is defined as
\begin{subequations}
\begin{align}
d_i &:= \lim_{P\to+\infty}\frac{R_i}{\log(1+\mathsf{P})}, \ i\in[1:2],
\\
\dof &:= d_1+d_2,
\label{eq:sum dof_def}
\end{align} 
for all possible achievable rate pairs $(R_1, R_2)$.
\label{eq:dof_def}
\end{subequations}

The main contribution of this paper is: 
\begin{thm}
\label{th:MainAch}
For the Gaussian MIMO two-user interference channel with an instantaneous relay, the largest achievable DoF region with a  linear strategy at the relay is
\begin{subequations}
\begin{align}
	d_1 &\leq  M,\\
	d_2 &\leq  M,\\
	d_1+d_2 &\leq  \dof^{\rm lin},
\end{align}
\label{eq:achievable_region}
\end{subequations}
where 
\begin{align}
\dof^{\rm lin}:= 
M+\min\left\{M,N,L,\frac{\max\left\{N,L\right\}}{2}\right\}.
\label{eq:achievability_sum}
\end{align}
Moreover, since any achievable DoF must satisfy
\begin{align}
\dof^{\rm lin} \leq \dof \leq M+\min\left\{M,N,L\right\},
\label{eq:it_converse}
\end{align}
$\dof^{\rm lin}$ is optimal whenever 
$\max\left\{N,L\right\} \geq 2\min\left\{M,N,L\right\}$.
\end{thm}

\begin{figure}
\centering
\includegraphics[width=0.9\columnwidth]{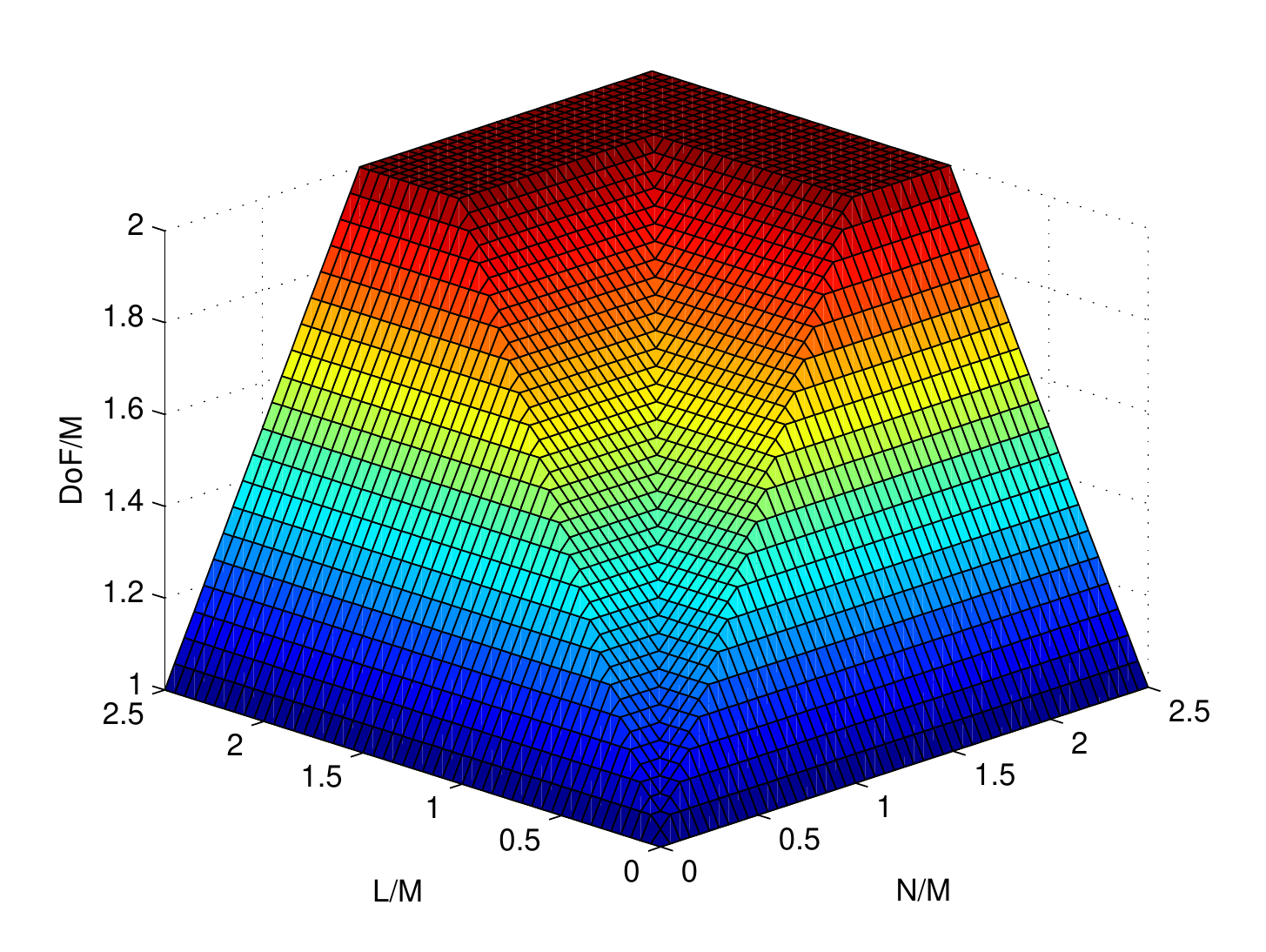}
\caption{Corresponding sum DoF for different settings of $(N,L)$.}
\label{fig:dof_region}
\end{figure}
Fig. \ref{fig:dof_region} shows $\dof^{\rm lin}/M$ as a function of $N/M$ and $L/M$. 
We observe that for specific $L$, increasing $N$ may not help to increase the sum DoF for certain regimes, namely $L/2<N<L$ and $2L<N$, and vice versa. 
Moreover, the shape of  $\dof^{\rm lin}/M$ is concave, meaning that if the limitation is on $N+L$, even distribution of the number of antennas is strictly suboptimal for sum DoF.

In Section~\ref{sec:achievability} we report the proof of the achievability of Theorem~\ref{th:MainAch}
and in Section~\ref{sec:converse} the converse.

\section{Achievability for Theorem~\ref{th:MainAch}} 
\label{sec:achievability}
We restrict attention to memoryless linear strategies at all nodes.
Let $d_{i}$ be the DoF achieved by user $i\in[1:2]$,
i.e., transmitter~$i$ transmits $d_{i}$ independent message streams.
The signals of interest are, for $i\in[1:2]$, as follows
\begin{subequations}
\begin{align}
X_0 &:= A Y_0 \in\mathbb{C}^{L \times 1}, 
\ \text{for some} \ 
A\in\mathbb{C}^{L \times N},
\label{eq:matAdef}
\\
X_i &:= V_i S_i\in\mathbb{C}^{M \times 1}, 
\\
V_i &:= [v_{i1},\ldots,v_{id_{i}}] \in\mathbb{C}^{M \times d_{i}},
\notag\\&
\ \text{for some} \ 
v_{ij} \in\mathbb{C}^{M \times 1},\ j\in[1:d_i],
\label{eq:matVidef}
\\
S_i &:= [x_{i1},\ldots,x_{id_{i}}]^T  \in\mathbb{C}^{d_{i} \times 1},
\\
T_i &:= A H_{0i} V_i = [t_{i1},\ldots,t_{id_{i}}] 
\notag\\&
\ \text{where} \ 
t_{ij} \in\mathbb{C}^{L \times 1},\ j\in[1:d_i],
\label{eq:matTidef}
\\
G &:= [H_{01}V_1 \  H_{02}V_2] = [g_{11} ,\ldots,g_{1d_1} \ g_{21}\ldots g_{2d_2}] 
\notag\\&
 \ \text{where} \ 
 g_{ij} \in\mathbb{C}^{N \times 1},\ j\in[1:d_i], 
\label{eq:matGdef}
\end{align} 
\end{subequations}
so that we can express the channel outputs as
\begin{subequations}
\begin{align}
Y_{0} &= 
  \sum_{d\in[1:d_{1}]}g_{1d}x_{1d}
 +\sum_{d\in[1:d_{2}]}g_{2d}x_{2d}+Z_0, 
\label{eq:Y_r}
\\
Y_{j} & =  Z_j+H_{j0}AY_0
 +\sum_{d\in[1:d_{1}]}\left(H_{j1}v_{1d}+H_{j0}t_{1d}\right)x_{1d}
\notag\\&\quad
 +\sum_{d\in[1:d_{2}]}\left(H_{j2}v_{2d}+H_{j0}t_{2d}\right)x_{2d},
  \ j\in[1:2],
\end{align}
\end{subequations}
where $x_{id}$ is the $d$-th independent message sent by transmitter $i$
in the direction of $v_{id}$, and
relayed by the relay in the direction of  $t_{id}=AH_{0i}v_{id}$, for $d\in[1:d_i]$ and $i\in[1:2]$.

We first give an intuition of the achievability of~\eqref{eq:achievable_region} in Section \ref{sub:intuition} to highlight two regimes of operation. In Section \ref{sub:examples} we provide examples for these two cases. In Section \ref{sub:case1} we proof the achievability for $L>N$ and in Section \ref{sub:case2} for $L\leq N$. We conclude with a discussion in Section~\ref{sub:achi_dicussion}.

\subsection{Intuition}
\label{sub:intuition}
We start by providing an intuition for the achievable DoF region in Theorem~\ref{th:MainAch}.
The aim of the relay is to {\it neutralize / cancel} the interference by 
choosing beamforming vectors that make
    $H_{10}t_{2d}+H_{12}v_{2d}=0$ (for receiver 1)
and $H_{20}t_{1d}+H_{21}v_{1d}=0$ (for receiver 2)
for as many messages as possible.
Since not all interfering messages can be neutralized, 
the interference is divided into two groups:
messages that are neutralized and those that are not;
the latter will be decoded at the non-intended receiver.
In our scheme, messages may be aligned at the relay
but no IA will occur at the receivers.

Let us start by ``counting'' how many messages could possibly be neutralized.
We distinguish two cases:

\paragraph{Case 1} 
Without message alignment at the relay, 
the relay can decode at most $N$ different linear combinations of the transmitted signals (since it has $N$ receiving antennas);
thus the number of neutralizations is limited by $N$. 
Consider next the solution of the equations $H_{10}t_{2d}+H_{12}v_{2d}=0$ and $H_{20}t_{1d}+H_{21}v_{1d}=0$; 
there can be at most $L$ linearly independent $t_{1d}$ and $L$ linearly independent $t_{2d}$;
thus $2L$ is another limitation for the number of neutralizations.
The number of neutralized messages is therefore limited by $\min\left\{ N,2L \right\}$.

\paragraph{Case 2} 
Consider now message alignment at the relay.  
The number of aligned message pairs at the relay can not exceed $N$. 
With alignment, the relay decodes $x_{1d}+x_{2d}$, and chooses $t_{1d}=t_{2d}=:t_{d}$ so that interference is neutralized at both receivers, which amounts to solving
\begin{align}
\begin{bmatrix}
H_{20} & H_{21} & 0\\
H_{10} & 0 & H_{12}\\
0 & H_{01} & -H_{02}
\end{bmatrix}
\begin{bmatrix}
t_{d}\\
v_{1d}\\
v_{2d}
\end{bmatrix}=0.
\label{eq:align_at_relay}
\end{align}
Since the `channel' matrix in~\eqref{eq:align_at_relay} has dimension $\left(2M+N\right) \times \left(2M+L\right)$,
the equation in~\eqref{eq:align_at_relay} 
has at least $[L-N]^+$ linearly independent solutions.
If $L>N$, the relay can at least neutralize $L-N$ message pairs simultaneously
at the receivers; the relay can further 
decode $N-(L-N)=2N-L$ linear combinations of the remaining messages,
which can be neutralized with the scheme discussed in Case~1;
the number of neutralizations for $L>N$ is thus limited by
$\min\left\{ 2N,2\left(L-N\right)+(2N-L)\right\}=\min\left\{ 2N,L\right\}$. 

\paragraph{Summary} 
\label{par:summary}
Having counted how many messages can be neutralized, the achievable DoF can be obtained as
$\mbox{\emph{\# antennas}}-(\mbox{\emph{\# interferers}}-\mbox{\emph{\# neutralizations}})$.
Let $D:=d_1+d_2$  to be the achievable sum DoF. 
For Case~1 we have 
$D = 2M-(D-\min\left\{ N,2L \right\})$, 
that is
$D = M+\min\left\{ \frac{N}{2},L \right\}$;
similarly for Case~2 we have 
$D = M+\min\left\{ \frac{L}{2},N \right\}$.
The intuition used to derive these two bounds also highlights the main ingredients of our achievable scheme for different antenna configurations, namely interference neutralization by the relay and message  alignment at the relay. 


\subsection{Examples} 
\label{sub:examples}
We show two examples to illustrate the intuitive arguments in Section~\ref{sub:intuition}.

\subsubsection{Case 1}
\label{par:case1}
Consider the case $(M,N,L)=(4, 4, 2)$. 
	
We show the achievability of the DoF region
\begin{align*}
		d_1 &\leq 4,\\
		d_1 &\leq 4,\\
		d_1+d_2 &\leq 6.
\end{align*}
The achievability is proved by achieving the DoF pairs $(d_1,d_2)=(4,2)$ and $(d_1, d_2)=(2,4)$. 
The received signal at the relay in \eqref{eq:channeloutputs} can be rewritten as
\begin{align*}
	Y_0 &= H_{01}V_1S_1+H_{02}V_2S_2+Z_0\\
	& = \begin{bmatrix}
		H_{02}V_2 & H_{01}V_1
	\end{bmatrix}
	\begin{bmatrix}
		S_2 \\ S_1
	\end{bmatrix}+Z_0,
\end{align*}
where $V_1\in \mathbb{C}^{4\times 4}$ and $V_2\in\mathbb{C}^{4\times 2}$. As long as the beamforming matrices $V_1$ and $V_2$ are not aligned at the relay, we are able to do Gaussian elimination for $\begin{bmatrix}
H_{02}V_2 & H_{01}V_1 	
\end{bmatrix}$, i.e., find a $U\in \mathbb{C}^{4\times 4}$ such that the first four columns of $U\begin{bmatrix}
H_{02}V_2 & H_{01}V_1 	
\end{bmatrix}$ form a diagonal matrix. We will show later that such a $U$ is guaranteed to exist almost surely.
After Gaussian elimination we have 
\begin{align}
	UY_0 &= [q_{11}, q_{12}, q_{21}, q_{22}]^T,\label{eq:gaussian_elimination}\\
 &\text{where}\nonumber\\
	q_{ij} &=  x_{ij}+l_{ij}(x_{13},x_{14}),\ j\in [1:d_i],\ i\in[1:2],\nonumber
\end{align}
where $l_{ij}(.,.)$ represents a the linear combination of its arguments. 
The relay sends 
\begin{align*}
	X_0 = TUY_0,
\end{align*}
where $T=[t_{11},t_{12},t_{21},t_{22}]\in\mathbb{C}^{2\times 4}$. 
The beamforming matrices $V_i, T_i, i\in[1:2]$, are chosen such that 
\begin{align}
	\begin{bmatrix}
		H_{i0} & H_{ij}
	\end{bmatrix}
	\begin{bmatrix}
		T_{j}\\ V_j
	\end{bmatrix}
	=0,\ i\neq j\in[1:2].
\end{align}
Since $\begin{bmatrix}
		H_{i0} & H_{ij}
	\end{bmatrix}$ has size $4 \times 6$, the first two columns of $V_1$ and $T_1$ can be chosen in the null space of  $\begin{bmatrix}
		H_{20} & H_{21}
	\end{bmatrix}$. Similarly,  $V_2, T_2$ can be chosen in the null space of  $\begin{bmatrix}
		H_{10} & H_{12}
	\end{bmatrix}$.
By construction, the beamforming matrices only depend $H_{12}, H_{10}, H_{21}, H_{20}$, i.e., they are independent of $H_{01},H_{02},H_{11},H_{22}$. By the generic channel assumption, this implies that the beamforming matrices are independent of the channel matrices $H_{01}, H_{02}$. Such independence shows that the matrix 
$\begin{bmatrix}
H_{02}V_2 & H_{01}V_1 	
\end{bmatrix}$ 
is full rank almost surely. 
Therefore,  the $U$ in \eqref{eq:gaussian_elimination} is guaranteed to exist.
By the choice of beamforming matrices, interfering messages $x_{21}$ and $x_{22}$ can be neutralized at receiver~1, thus making receiver~1 interference free. receiver~1 is therefore able to decode all four desired messages, namely $x_{11},x_{12},x_{13},x_{14}$. 
Interfering messages $x_{11}$ and $x_{12}$ are similarly neutralized  at receiver~2; the remaining interference messages $x_{13}$ and $x_{14}$ occupy  a two dimensional space at receiver~2. Since there are four antennas at the receiver~2, two desired messages can be decoded at receiver~2, together with two unintended messages from transmitter~1.
Therefore the proposed scheme achieves $(d_1,d_2)=(4,2)$. By symmetry the same scheme achieves $(d_1,d_2)=(2,4)$ too. 
Time sharing among these corner points concludes the proof of the claimed achievable region.

\subsubsection{Case 2} 
\label{par:case_2}
Now consider the case $(M,N,L)=(4, 2, 4)$. We prove that the DoF region
\begin{align*}
		d_1 &\leq 4,\\
		d_1 &\leq 4,\\
		d_1+d_2 &\leq 6,
\end{align*}
is achievable by achieving DoF pairs $(d_1, d_2)= (4,2)$ and $(d_1, d_2)= (2,4)$.
Following the intuition for Case 2 in Section~\ref{sub:intuition}, we solve \eqref{eq:align_at_relay}.
The null space of the matrix
\begin{align*}
	\begin{bmatrix}
H_{20} & H_{21} & 0\\
H_{10} & 0 & H_{12}\\
0 & H_{01} & -H_{02}
\end{bmatrix}\in\mathbb{C}^{10 \times 12}
\end{align*}
has dimension at least two, so we can take  
$\begin{bmatrix}
T &
V_{1} &
V_{2} \\
\end{bmatrix}^T$ as two columns in its null space. Then, by \eqref{eq:matTidef}, 
the relay processing matrix is
$A = T(H_{01}V_1)^{-1}$.
Since $V_1$ is independent of $H_{01}$ and $H_{01}V_1$ is full rank, we have that  $(H_{01}V_1)^{-1}$ exists almost surely. Thus the relay can find an $A$ such that the acheivability scheme neutralizes two interfering messages at each receiver. 
Since receiver~1 has a four dimensional interference free space and receiver~2 has a  two dimensional interference free space, the DoF pair $(d_1, d_2)=(4,2)$ is achievable. 
By symmetry, DoF pair $(d_1, d_2)=(2,4)$ is achieved by the same scheme.
By time sharing among these corner points the claimed region is achievable.

\subsubsection{Discussion}
Note that the scheme in~\cite{2ic-ir} for the case $M=N=L=4$ achieves a sum DoF of 6. In our two examples we achieve the same DoF but with either less antennas at the relay receiving side or less antennas at the relay transmitting side. This shows that some antennas may be redundant in terms of DoF.


\subsection{Achievability for Case 1: $L\leq N$}
\label{sub:case1}
We assume parameters $N$ and $L$ to be even whenever necessary; if not, the same strategy applies over two channel uses. 
%
Alignment at the relay is not possible in this case because the equation in~\eqref{eq:align_at_relay} has no solutions
almost surely for generic channel matrices. We show that the DoF pairs $(M,\min\{L,M,N/2\})$ and $(\min\{L,M,N/2\},M)$ are achievable.
Let $d_1=M$, $d_2=\min\{L,M,N/2\}$, 
and let $K=\min\{L,M,N/2\}$ represent the number of interfering messages at each receiver that are neutralized thanks to the relay. 
Indeed, the constraint $2K\leq N$ guarantees that the relay can neutralize $K$ messages per user,
$K\leq L$ that neutralization is possible by the relay alone, and
$K\leq M$ that a certain transformation at the relay is possible (i.e., to find the matrix indicated by $U$ in the following).

The relay receives $Y_{0}= G  [S_1^T \ S_2^T]^T + Z_0$ as in~\eqref{eq:Y_r},
and decodes $2K$ linear combinations of messages.
Since $v_{id}\neq 0$ (otherwise the corresponding message is not sent)
and $H_{0i}$ is generic, $g_{id}=H_{0i}v_{id}\neq 0$ almost surely. 
Moreover, the matrix $G$ in (\ref{eq:matGdef}) is full rank  almost surely, 
so we can find a full-rank matrix $U\in\mathbb{C}^{2K \times N}$ 
(note $2K\leq\rk(G)=\min\{N,D\}$) such that 
\begin{align*}
&UY_{0} =
\begin{bmatrix}
q_{11},\ldots,q_{1K} q_{21},\ldots,q_{2K}
\end{bmatrix}^T, \ \text{where}
\\
&q_{1j}  =  x_{1j}+l_{j}
\left(x_{1\left(K+1\right)},\ldots,x_{1 d_1},x_{2\left(K+1\right)},\ldots,x_{2 d_2}\right), 
\\ 
&q_{2j}  =  x_{2j}+l_{K+j}
\left(x_{1\left(K+1\right)},\ldots,x_{1 d_1},x_{2\left(K+1\right)},\ldots,x_{2d_2}\right), 
\end{align*}
for $j\in[1:K],$ where each element of $UY_{0}$ contains one of the messages
to be neutralized (here $x_{id}, \ i\in[1:2], \ d\in[1:K]$) and a
linear combination (here $l_{j}\left(\ldots\right), \ j\in[1:2K]$) of the remaining messages.
The relay sends 
\begin{align*}
	X_0& = \sum_{i=1}^{K}  t_{1i} q_{1i}+\sum_{i=1}^{K} t_{2i} q_{2i}\\
	&=TUY_0=AY_0,
\end{align*}
where $T=[t_{1i},\ldots t_{1K},t_{21},\ldots,t_{2K}]$.
Next, neutralization of the interference is done separately at each receiver
by solving
\begin{align}
\begin{bmatrix}
  H_{i0} & H_{ij}\\
\end{bmatrix}
\begin{bmatrix}
t_{jd}\\
v_{jd}
\end{bmatrix} &=0
,
\begin{array}{r}
i\not=j\in[1:2],\\
d\in[1:K],\\
\end{array}
\label{eq:INconditions}
\end{align}
that is, we seek $K$ column vectors 
$\left[\begin{smallmatrix}
t_{jd}\\
v_{jd}
\end{smallmatrix}\right]\in \mathbb{C}^{(L+M) \times 1}$
in the null space of 
$\left[H_{i0}\quad H_{ij}\right]\in \mathbb{C}^{M \times (L+M)},$ 
which has dimension at least $L$.
Since $K\leq L$, such vectors can be found almost surely.
Moreover, 
the columns in null spaces are linearly
independent, i.e., $v_{1d}$ and $v_{2d}$ are linearly independent,
and thus also  
$g_{id}=H_{0i}v_{id}$
are linearly independent (because they are random projections of $v_{id}$), which shows that $G$ in (\ref{eq:matGdef}) is full rank almost surely.
%
With $K$ interfering messages neutralized at each receiver, receiver~1 becomes interference free, while receiver~2 is left with a $K$ dimension interference free space thus it is able  to decode $K=\min\{L,M,N/2\}$ messages. 
This shows the achievability of the DoF pair $(M, \min\{L,M,N/2\})$.
Similarly it can be shown that the DoF pair $(\min\{L,M,N/2\}, M)$ is also achievable. 
Then, by time sharing, the DoF region
\begin{subequations}
\begin{align}
	d_1 &\leq M,\\
	d_2 &\leq M,\\
	d_1+d_2 &\leq M+\min\{L,M,N/2\},
\end{align} 
is achievable for the case $L\leq N$. 
\label{eq:dof_region_L_less_N}
\end{subequations}

\subsection{Achievability for Case 2: $L> N$}
\label{sub:case2}
Again We assume parameters $N$ and $L$ to be even whenever necessary; if not, the same strategy applies over two channel uses. 
The users send $d_1=M$, $d_2=\min\{L/2,M,N\}$ independent messages. 
In this case, the equation in~\eqref{eq:align_at_relay}
has 
solutions so alignment at the relay is possible. 
As discussed in Section \ref{sub:intuition}, there will be two kinds of neutralizations:
messages that are aligned at the relay and neutralized, 
and those that are not aligned at the relay and dealt with by applying the scheme in Section~\ref{sub:case1}.
Let $K=\min\{L/2,M,N\}$ represent the number of interfering messages which are neutralized at each receiver. We divide $K=K_1+K_2$, where $K_1$ is the number of interfering messages that are aligned at the relay, and $K_2$ is the number of interfering messages that are not aligned. Based on \eqref{eq:align_at_relay} we can have $K_1=\min\{N,L-N,M\}$ messages aligned at the relay. 
Therefore the relay receives $Y_0=G[S_1^T S_2^T]+Z_0$ and decodes $K_1+2K_2$ linear combinations of messages. Since we have $H_{01}v_{1d}=H_{02}v_{2d},d\in[1:K_1]$, we can find a full-rank matrix $U\in\mathbb{C}^{K_1+2K_2 \times N}$ such that 
\begin{align*}
UY_{0} &=
\begin{bmatrix}
q_{11},\ldots,q_{1K_1} q_{21},\ldots,q_{2K_2} q_{31},\ldots,q_{3K_2} 
\end{bmatrix}^T, \ \text{where}
\\
q_{1i}  &=  x_{1i}+x_{2i}+l_{i}
\left(x_{1\left(K+1\right)},\ldots,x_{1 d_1},x_{2\left(K+1\right)},\ldots,x_{2d_2}\right), 
\\ &\text{for } i\in[1:K_1]
\\ 
q_{2j}  &=  x_{2j}+l_{K_1+j},
\left(x_{1\left(K+1\right)},\ldots,x_{1 d_1},x_{2\left(K+1\right)},\ldots,x_{2d_2}\right),
\\
q_{3j}  &=  x_{3j}+l_{K_1+K_2+j}
\left(x_{1\left(K+1\right)},\ldots,x_{1 d_1},x_{2\left(K+1\right)},\ldots,x_{2d_2}\right),\\
&\text{for } j\in[1:K_2].
\end{align*}
The relay sends 
\begin{align*}
	X_0 &= \sum_{i=1}^{K_1}  t_{i} q_{1i}+\sum_{i=1}^{K_2} t_{1i} q_{2i} +\sum_{i=1}^{K_2}t_{2i}q_{3i}\\
	&=TUY_0=AY_0.
\end{align*}
In this case the vectors $v_{1d}, v_{2d}, t_d, d\in[1:K_1],$ are chosen from the solutions of~\eqref{eq:align_at_relay}. Following the generic channel setting, such vectors
exist almost surely.
This would make $K_1$ interfering messages neutralized at each receiver. 
The choice of $v_{jd},j\in[1:2],d\in[K_1:d_j]$, $t_{jd}, j\in[1:2],d\in[1:K_2],$ follows the same argument in Section \ref{sub:case1}, namely, $v_{jd},j\in[1:2],d\in[K_1:d_j],$ and $t_{jd}, j\in[1:2],d\in[1:K_2],$ are such that
\begin{align*}
\begin{bmatrix}
  H_{i0} & H_{ij}\\
\end{bmatrix}
\begin{bmatrix}
t_{jd}\\
v_{j(K_1+d)}
\end{bmatrix} &=0
,
\begin{array}{r}
i\not=j\in[1:2],\\
d\in[1:K_2].\\
\end{array}
\end{align*} 

By the proposed scheme, decoding is possible if 
\begin{align*}
   K_1     &\leq [L-N]^+,  & \text{for alignment at the relay,}
\\ K_1+2K_2&\leq N,        & \text{decoding at the relay,} 
\\      K_2&\leq L,        & \text{neutralization by the relay,}
\\ d_1+d_2-K_1-K_2&\leq M, & \text{decoding at the receivers,}
\end{align*}
thus 
in order to maximize the sum DoF $d_1+d_2= K_1+K_2+M$ we solve
\begin{align*}
   \max\{K_1+K_2\} &: \text{subject to}
\\ K_1&\leq\min([L-N]^+,N,M)  
\\ K_2&\leq\min(N/2,L,M)       
\\ K_1+ K_2&\leq M 
\\ K_1+2K_2&\leq N. 
\end{align*}
The optimal solution can be easily found to be
\begin{align*}
 K=K_1+K_2 &= \min\{L/2,N,M\}.
\end{align*}
This shows that a total of $K=\min\{L/2,M,N\}$ interfering messages per users can be neutralized at the non-intended receiver. 
receiver~1 sees no interference, therefore it can  decode $M$ messages, while receiver~2 has a $\min\{L/2,M,N\}$ dimensional interference free space, thus can decode $\min\{L/2,M,N\}$ messages. This scheme shows that the DoF pair $(d_1,d_2) = (M, \min\{L/2,M,N\})$ is achievable. By symmetry $(d_1,d_2) = (\min\{L/2,M,N\}, M)$ is also achievable by the same strategy.
Again, by time sharing, the DoF region
\begin{subequations}
\begin{align}
	d_1 &\leq M,\\
	d_2 &\leq M,\\
	d_1+d_2 &\leq M+\min\{L/2,M,N\},
\end{align}
is achievable for the case $L> N$.
\label{eq:dof_region_L_more_N}
\end{subequations}

Combining \eqref{eq:dof_region_L_less_N} and \eqref{eq:dof_region_L_more_N} concludes the proof of achievability.

\subsection{Discussion} 
\label{sub:achi_dicussion}
In our scheme message alignment only occurs at the relay. An interesting question is whether forcing interfering messages to align at the non-intended receiver, as in IA, would further increase the DoF. Interestingly, for the considered 2IC+IR it can be shown that IA does not have an advantage over interference neutralization at the receivers in terms of DoF. 

Consider doing IA at receiver~1. Without loss
of generality, assume we are going to align $x_{21}$ into the space
of $x_{2d},d\in I$, where $I$ is the set of all interfering messages that
have not been aligned or neutralized yet. 
The subspace spanned by $I$ is available for IA. As in the above
proof, let $v_{2d}$  be the beamforming vector for $x_{2d}$ chosen by the
transmitter and $t_{2d}$ be the beamforming vector at relay for $x_{2d}$. Then we impose
\begin{align*}
H_{12}v_{21}+H_{10}t_{21}\in\mbox{span}\left\{ H_{12}v_{2d}+H_{10}t_{2d} ,d\in I\right\},
\end{align*}
that is
\begin{align*}
\begin{bmatrix}
H_{12} & H_{10}\end{bmatrix}\left(\begin{bmatrix}
v_{21}\\
t_{21}
\end{bmatrix}-\sum_{d}a_{d}\begin{bmatrix}
v_{2d}\\
t_{2d}
\end{bmatrix}\right)=0,
\end{align*}
for some real number $a_d,d\in I$.
In other words, $v_{2d}$ and $v_{21}$ must be linearly
independent and $\begin{bmatrix}
v_{21}\\
t_{21}
\end{bmatrix}-\sum_{d}a_{d}\begin{bmatrix}
v_{2d}\\
t_{2d}
\end{bmatrix}$ must be a linear combination of the columns in the null space of $\begin{bmatrix}
H_{12} & H_{10}\end{bmatrix}$.

Then consider the alignment of $x_{22}$. 
By the same reasoning, $v_{2d}$ and $v_{22}$ must be linearly independent and 
$\begin{bmatrix}
v_{22}\\
t_{22}
\end{bmatrix}-\sum_{d}b_{d}\begin{bmatrix}
v_{2d}\\
t_{2d}
\end{bmatrix}$ must be a linear combination of the columns of the null space of $\begin{bmatrix}
H_{12} & H_{10}\end{bmatrix}$, for some real number $b_d,d\in I$.

Noting that $v_{21}$, $v_{22}$ and $v_{2d}$  also have to
be linearly independent, the linear combinations
we chose for $v_{21}$ and $v_{22}$ must be linearly independent too.
This indicates that the number
interfering messages that we can align is limited by the dimension of the null space of the matrix
$\begin{bmatrix}
H_{12} & H_{10}\end{bmatrix}$. Interestingly, this is the same condition for interference neutralization in \eqref{eq:INconditions}.
This indicates that for every interfering message that can be aligned, there is enough
``linear independence'' for its neutralization.

Therefore, this shows that the constraints that IA imposes are ultimately the same as for interference neutralization in~\eqref{eq:INconditions}.

\section{Converse for Theorem~\ref{th:MainAch}} 
\label{sec:converse}

Trivially we have upper bounds $d_i\leq M, i\in[1:2],$ from interference free MIMO channels. Thus, in order to prove the region in \eqref{eq:achievable_region}, we only need to show the sum DoF bound in \eqref{eq:achievability_sum}.
In this section we provide two types of converse proofs for the sum DoF in Theorem~ \ref{th:MainAch}.

\subsection{Genie-aided Information Theoretical Upper Bounds} 
\label{sub:information_theoretical_upper_bounds}

\subsubsection{Cognitive Bound} 
\label{ssub:cognitive_bound}

If one transmitter and the relay perfectly cooperate and in addition are given the message of the other transmitter, we obtain a two-user cognitive interference channel, with $M+L$ antennas at the cognitive transmitter, $M$ antennas at the primary transmitter and $M$ antennas at each receiver, whose DoF can be evaluated from the constant gap result in~\cite[Th.V.3]{rini:JSAC2014}; thus
\begin{align}
  \dof\leq \min\{M+L, 2M\}.
\label{eq:cognitive_bound}
\end{align}

\subsubsection{Genie Bound} 
\label{ssub:genie_bound}

By providing the received signal at the relay to both receivers we obtain a classical $2$-user IC with $M$ antennas at each transmitter and $M+N$ antennas at each receiver whose DoF was found in~\cite[eq.(3)]{rk-deficient-mimo}; thus
\begin{align}
  \dof\leq \min\{M+N, 2M\}.
\label{eq:genie_bound}
\end{align}
The bounds in~\eqref{eq:cognitive_bound}-\eqref{eq:genie_bound} prove~\eqref{eq:it_converse} and that 
$\dof=\dof^{\rm lin} = M+\min\{M,N,L\}$  for  $\max\left\{N,L\right\} \geq 2\min\left\{M,N,L\right\}$.



\subsection{Optimality Among Linear Schemes with Symbol-by-Symbol Processing} 
\label{sub:linear_optimality}
At the time of submitting this manuscript, an information theoretic converse for 
the regime $\max\left\{N,L\right\} < 2\min\left\{M,N,L\right\}$,
for which $\dof^{\rm lin} = M+\max\{N,L\}/2$, was not available.
In the rest of the section we show that $\dof^{\rm lin}$
in~\eqref{eq:achievability_sum} is the largest possible DoF when the relay is restricted to linear operations.
Note that no restrictions are imposed on the operations at the transmitters.

Before giving the proof details, 
in Section~\ref{ssub:matrix_multiplication_and_schur_complement}
we introduce a property of the rank of a partitioned matrix which will be useful when 
dealing with the partitioned matrices introduced in Section~\ref{ssub:channel_transformation_based_on_channel_coefficient_matrix}.
The proof of optimality of our scheme among linear schemes at the relay is then divided in different cases and analyzed in Sections~\ref{ssub:case n=l=m} to~\ref{sub:case_n_neq_l}.

\subsubsection{Matrix multiplication and Schur complement} 
\label{ssub:matrix_multiplication_and_schur_complement}

In the following, the multiplication of matrices $M_{1}\in \mathbb{C}^{l_{1}\times l_{2}}$ and
$M_{2}\in \mathbb{C}^{l_{2}\times l_{3}}$ is denoted as $M_{1}M_{2}\in \mathbb{C}^{l_{1}\times l_{3}}$.
When either $l_{1}=0$ or $l_{3}=0$,
the matrix $M_{1}M_{2}$ has dimension zero, i.e., it does not exist. When
$l_{1}\neq0, l_{3}\neq0$, and $l_{2}=0$, the matrix $M_{1}M_{2}$
is defined as the all zero matrix of dimension $l_{1}\times l_{3}$. 

We will use the Schur complement and its property:
for a partitioned square matrix 
$M=\left[\begin{smallmatrix}
m^{11} & m^{12}\\
m^{21} & m^{22}
\end{smallmatrix}\right]$ where $m^{22}$ is square and nonsingular, the Schur complement
of $m^{22}$
is defined as $M/m^{22}:=m^{11}-m^{12}\left(m^{22}\right)^{-1}m^{21}$.
By \emph{Guttman rank additivity formula} \cite{guttman_rank_additivity} we have
\begin{align}
\rk\left(M\right)=\rk\left(m^{22}\right)+\rk\left(M/m^{22}\right).
\label{eq:Schurrkadd}
\end{align}

\subsubsection{Channel Transformation} 
\label{ssub:channel_transformation_based_on_channel_coefficient_matrix}

We partition the encoding matrix used at the relay as
\begin{align}
A:=\begin{bmatrix}
a^{11} & a^{12}\\
a^{21} & a^{22}
\end{bmatrix}
:
\begin{array}{l}
a^{11}\in \mathbb{C}^{\min\left\{ L,M\right\} \times\min\left\{ N,M\right\}},\\
a^{12}\in \mathbb{C}^{\min\left\{ L,M\right\} \times\left(N-M\right)^{+}},\\
a^{21}\in \mathbb{C}^{\left(L-M\right)^{+}\times\min\left\{ N,M\right\}},\\
a^{22}\in \mathbb{C}^{\left(L-M\right)^{+}\times\left(N-M\right)^{+}}.
\end{array}
\label{eq:Aprtinioned}
\end{align}
Similarly, we partition the channel matrices as
\begin{align}
H_{j0} & :=  \begin{bmatrix}
h_{j0}^{11} & h_{j0}^{12} \\
\end{bmatrix} :
\begin{array}{l}
h_{j0}^{11}\in \mathbb{C}^{M\times\min\left\{ L,M\right\}},\\
h_{j0}^{12}\in \mathbb{C}^{M\times\left(L-M\right)^{+}},\\
\end{array}
\\
H_{0j} & :=  \begin{bmatrix}
h_{0j}^{11}\\
h_{0j}^{21} \\
\end{bmatrix}:
\begin{array}{l}
h_{0j}^{11}\in \mathbb{C}^{\min\left\{ N,M\right\} \times M},\\
h_{0j}^{21}\in \mathbb{C}^{\left(N-M\right)^{+}\times M},\\
\end{array} 
\end{align}
and  define

\begin{align}
H_{j0}^{\prime} & :=  \begin{bmatrix}
\left(h_{j0}^{11}\right)^{\dagger}\\
\nul\left(h_{j0}^{11}\right) \\
\end{bmatrix}:
\begin{array}{l}
\left(h_{j0}^{11}\right)^{\dagger}\in \mathbb{C}^{\min\left\{ L,M\right\} \times M},\\
\nul\left(h_{j0}^{11}\right)      \in \mathbb{C}^{\left(M-L\right)^{+}\times M},\\
\end{array} 
\label{eq:Hj0_prime}\\
H_{0j}^{\prime} & :=  \begin{bmatrix}
\left(h_{0j}^{11}\right)^{\dagger} & \nul\left(h_{0j}^{11}\right) \\
\end{bmatrix}:
\begin{array}{l}
\left(h_{0j}^{11}\right)^{\dagger}\in \mathbb{C}^{M\times \min\left\{ N,M\right\}},\\
\nul\left(h_{0j}^{11}\right)      \in \mathbb{C}^{M\times\left(M-N\right)^{+}},\\
\end{array} 
\label{eq:H0j_prime}
\end{align}
for $j\in[1:2]$.
Note that some matrices will have dimension zero under certain settings. 

The matrices $H_{j0}^{\prime}$	 in \eqref{eq:Hj0_prime} and $H_{0j}^{\prime}$  in \eqref{eq:H0j_prime} are full-rank square matrices for $j\in[1:2]$,
owing to the fact that all channel matrices are generic and independent. 
%
Hence the invertible transformation 
\begin{align}
\widehat{Y}_{j} &=  H_{j0}^{\prime}Y_{j}, \nonumber\\
\widehat{X}_{j} &=  \left(H_{0j}^{\prime}\right)^{-1}X_{j}, \ \ j\in[1:2],
\label{eq:transformationwlog}
\end{align}
is without loss of optimality. 
The transformed channel outputs (neglecting the noise terms) can be written as
\begin{align}
\widehat{Y}_{1} 
&= \left(H_{10}^{\prime}H_{11}H_{01}^{\prime}+G_{11}\right)\widehat{X}_{1}
 + \left(H_{10}^{\prime}H_{12}H_{02}^{\prime}+G_{12}\right)\widehat{X}_{2},
\label{eq:transformationwlogy1}
\\
\widehat{Y}_{2} 
&= \left(H_{20}^{\prime}H_{21}H_{01}^{\prime}+G_{21}\right)\widehat{X}_{1}
 + \left(H_{20}^{\prime}H_{22}H_{02}^{\prime}+G_{22}\right)\widehat{X}_{2},
\label{eq:transformationwlogy2}
\end{align}
where the matrices $G_{ij}, (i,j)\in[1:2]^2$, defined in~\eqref{eq:def_of_G}  at the top of next page,
are the only ones that depend on $A$ in~\eqref{eq:Aprtinioned}.

The transformed channel in~\eqref{eq:transformationwlogy1}-\eqref{eq:transformationwlogy2} is a classical $2$-user MIMO IC, whose DoF
gives the DoF of our original channel when the relay is restricted to linear operations.
The sum DoF of the classical $2$-user MIMO IC was found in 
~\cite[eq.(3)]{rk-deficient-mimo}
under the assumption that all channel matrices were are the product of two generic matrices. 
The result in~\cite[eq.(3)]{rk-deficient-mimo} is not applicable in our case since the transformed channel matrices may have a special structure (as the all depend on the matrix $A$ chosen by the relay) which is not equivalent to the assumption leading to~\cite[eq.(3)]{rk-deficient-mimo}. Therefore in Appendix \ref{sec:appen_general-2ic} we derive the DoF region of the $2$-user MIMO IC with {\it arbitrary channel matrices} from the constant gap of \cite{IC_constant_gap}. This DoF region is of interest in itself.

The best linear scheme for our original channel is thus the one that optimizes the DoF of channel in~\eqref{eq:transformationwlogy1}-\eqref{eq:transformationwlogy2} over the choice of $A$ in~\eqref{eq:Aprtinioned}.
By considering the upper bounds in~\eqref{eq:general_dof_region d1+d2 kra1}-\eqref{eq:general_dof_region d1+d2 kra2}
and trivial upper bound on the terms `$+\rk(\cdot)$', we have
\begin{align*}
	d_1+d_2 &
		\leq 2M-\rk\left(H_{10}^{\prime}H_{12}H_{02}^{\prime}+G_{12}\right),
		\\
	d_1+d_2 &
		\leq 2M-\rk\left(H_{20}^{\prime}H_{21}H_{01}^{\prime}+G_{21}\right),
\end{align*}
that is, $\dof^{\rm lin}$ must satisfy~\eqref{eq:doflinupper}  at the top of next page.
We therefore proceed now to evaluate, or lower bound, the term $\Delta$ in~\eqref{eq:def_of_delta}
 at the top of next page in order to upper bound $\dof^{\rm lin}$.	
Next we analyze different antennas configurations separately.

\begin{figure*}[!t]
\begin{align}
	G_{ij}&=
	\begin{bmatrix}
	a^{11}+a^{12}h_{0j}^{21}\left(h_{0j}^{11}\right)^{\dagger}+\left(h_{i0}^{11}\right)^{\dagger}h_{i0}^{12}\left(a^{21}+a^{22}h_{0j}^{21}\left(h_{0j}^{11}\right)^{\dagger}\right) 
	& \left(a^{12}+\left(h_{i0}^{11}\right)^{\dagger}h_{i0}^{12}a^{22}h_{0j}^{21}\right)\nul\left(h_{0j}^{11}\right)
\\
  	\nul\left(h_{i0}^{11}\right)h_{i0}^{12}\left(a^{21}+a^{22}h_{0j}^{21}\left(h_{0j}^{11}\right)^{\dagger}\right) 
	& \nul\left(h_{i0}^{11}\right)h_{i0}^{12}a^{22}h_{0j}^{21}\nul\left(h_{0j}^{11}\right)
\end{bmatrix},
\label{eq:def_of_G}
\\
   \dof^{\rm lin} 
   &
   \leq 2M-\Delta,
\label{eq:doflinupper}
      \\
   \Delta &:= \min_{A}
   \frac{\rk\left(H_{10}^{\prime}H_{12}H_{02}^{\prime}+G_{12}\right)+\rk\left(H_{20}^{\prime}H_{21}H_{01}^{\prime}+G_{21}\right)}{2}.
\label{eq:def_of_delta}
\end{align}
\hrulefill
\vspace*{4pt}
\end{figure*}

\subsubsection{Case $N=L=M$} 
\label{ssub:case n=l=m}
In this case $(N-M)^+=(L-M)^+=0$, thus in \eqref{eq:doflinupper} we have $G_{11}=G_{12}=G_{21}=G_{22}=a^{11}$. 
Since $\rk(A)=\rk(-A)$, we have 
\begin{align*}
  & \quad    \rk\left(H_{10}^{\prime}H_{12}H_{02}^{\prime}+a^{11}\right)+\rk\left(-H_{20}^{\prime}H_{21}H_{01}^{\prime}-a^{11}\right)\\
  &\geq \rk\left(H_{10}^{\prime}H_{12}H_{02}^{\prime}+a^{11}-H_{20}^{\prime}H_{21}H_{01}^{\prime}-a^{11}\right)\\
  &=    \rk\left(H_{10}^{\prime}H_{12}H_{02}^{\prime}-H_{20}^{\prime}H_{21}H_{01}^{\prime}\right)\\
   &=  M,
\end{align*}
where the inequality follow from~\cite[eq.(0.4.5.d)]{Horn:1985:MA:5509},
%
%
which implies $\Delta \geq M$ and thus
 \begin{align}
  \dof^{\rm lin} 
   \leq  2M-M/2 =  3M/2.
   \label{eq:doflinupn=l=m}
 \end{align}
 The upper bound in~\eqref{eq:doflinupn=l=m} matches the achievable sum DoF in~\eqref{eq:achievability_sum}, thus showing the optimality of our scheme among all memoryless linear processing schemes at the relay.

\subsubsection{Case $N=L<M$} 
\label{ssub:case n=l<m}
In this case we have
%
%
\begin{align*}
G_{12}  =  
G_{21}  =  \begin{bmatrix}
a^{11} & \mathbf{0}_{N\times\left(M-N\right)}\\
\mathbf{0}_{\left(M-N\right)\times N} & \mathbf{0}_{\left(M-N\right)\times\left(M-N\right)}
\end{bmatrix}.
\end{align*}

Let
\begin{align*}
H_{i0}^{\prime}H_{ij}H_{0j}^{\prime}
&=\begin{bmatrix}
F_{ij}^{11} & F_{ij}^{12}\\
F_{ij}^{21} & F_{ij}^{22}\\
\end{bmatrix}:
\begin{array}{l}
F_{ij}^{11} \in \mathbb{C}^{N \times N},\\
F_{ij}^{12} \in \mathbb{C}^{N \times (M-N)},\\
F_{ij}^{21} \in \mathbb{C}^{(M-N) \times N},\\
F_{ij}^{22} \in \mathbb{C}^{\left(M-N\right) \times\left(M-N\right)},\\
\end{array}
\end{align*}
$i\not=j\in[1:2]$. Note that 
$F_{ij}^{22}=\nul\left(h_{i0}^{11}\right)H_{ij}\nul\left(h_{0j}^{11}\right)$
is a full-rank square matrix almost surely.
By applying~\eqref{eq:Schurrkadd} 
we have
\begin{align*}
  &  \rk\left(H_{10}^{\prime}H_{12}H_{02}^{\prime}+G_{12}\right)+\rk\left(H_{20}^{\prime}H_{21}H_{01}^{\prime}+G_{21}\right)
\\=& \rk\left(F_{12}^{22}\right)+\rk\left(\left(H_{10}^{\prime}H_{12}H_{02}^{\prime}+G_{12}\right)/F_{12}^{22}\right)
\\ &+\rk\left(F_{21}^{22}\right)+\rk\left(\left(H_{20}^{\prime}H_{21}H_{01}^{\prime}+G_{21}\right)/F_{21}^{22}\right)
\\= &\rk\left(F_{12}^{22}\right)+\rk\left(a^{11}+F_{12}^{11}-F_{12}^{12}\left(F_{12}^{22}\right)^{-1}F_{12}^{21}\right)
\\&+ \rk\left(F_{21}^{22}\right)+\rk\left(a^{11}+F_{21}^{11}-F_{21}^{12}\left(F_{21}^{22}\right)^{-1}F_{21}^{21}\right)
\\\geq & \rk\left(F_{12}^{22}\right)+\rk\left(F_{21}^{22}\right)
\\&+\rk\left(F_{12}^{11}-F_{12}^{12}\left(F_{12}^{22}\right)^{-1}F_{12}^{21}-F_{21}^{11}+F_{21}^{12}\left(F_{21}^{22}\right)^{-1}F_{21}^{21}\right)
\\= &2\left(M-N\right)+N\\
  =& 2M-N,
\end{align*}
where the matrix $F_{12}^{11}-F_{12}^{12}\left(F_{12}^{22}\right)^{-1}F_{12}^{21}-F_{21}^{11}+F_{21}^{12}\left(F_{21}^{22}\right)^{-1}F_{21}^{21}$ is full rank due to the fact that all the components of the summation are full rank almost surely and mutually independent.
This implies
 \begin{align}
   \dof^{\rm lin} 
   &\leq  2M-\frac{2M-N}{2}=  M+\frac{N}{2}.
   \label{eq:doflinupn=l<m}
 \end{align}
The upper bound in~\eqref{eq:doflinupn=l<m} matches the achievable sum DoF in~\eqref{eq:achievability_sum}, thus showing the optimality of our scheme among all memoryless linear processing schemes at the relay.

\subsubsection{Case $N=L>M$} 
\label{ssub:case n=l>m}
In this case we have
\begin{align*}
 G_{12} &=  a^{11}+a^{12}H_{02}^{21}\left(H_{02}^{11}\right)^{-1}\\
  &\quad +\left(H_{10}^{11}\right)^{-1}H_{10}^{12}\left(a^{21}+a^{22}H_{02}^{21}\left(H_{02}^{11}\right)^{-1}\right),\\
G_{21}  &=  a^{11}+a^{12}H_{01}^{21}\left(H_{01}^{11}\right)^{-1}\\
   &\quad+\left(H_{20}^{11}\right)^{-1}H_{20}^{12}\left(a^{21}+a^{22}H_{01}^{21}\left(H_{01}^{11}\right)^{-1}\right).
\end{align*}
We can bound the sum of the rank of the crosslinks by
\begin{align*}
   & \rk\left(H_{10}^{\prime}H_{12}H_{02}^{\prime}+G_{12}\right)+\rk\left(H_{20}^{\prime}H_{21}H_{01}^{\prime}+G_{21}\right)\\
  \geq & \rk\left(H_{10}^{\prime}H_{12}H_{02}^{\prime}+G_{12}-H_{20}^{\prime}H_{21}H_{01}^{\prime}-G_{21}\right)\\
  = & \rk(H_{10}^{\prime}H_{12}H_{02}^{\prime}-H_{20}^{\prime}H_{21}H_{01}^{\prime}\\
  & +a^{12}\left(H_{02}^{21}\left(H_{02}^{11}\right)^{-1}-H_{01}^{21}\left(H_{01}^{11}\right)^{-1}\right)\\
  & +\left(\left(H_{10}^{11}\right)^{-1}H_{10}^{12}-\left(H_{20}^{11}\right)^{-1}H_{20}^{12}\right)a^{21}\\
   & +\left(\left(H_{10}^{11}\right)^{-1}H_{10}^{12}-\left(H_{20}^{11}\right)^{-1}H_{20}^{12}\right)a^{22}H_{02}^{21}\left(H_{02}^{11}\right)^{-1}\\
  & +\left(H_{20}^{11}\right)^{-1}H_{20}^{12}a^{22}\left(H_{02}^{21}\left(H_{02}^{11}\right)^{-1}-H_{01}^{21}\left(H_{01}^{11}\right)^{-1}\right)).
\end{align*}
Define now two matrices
\begin{align}
R &=  \begin{bmatrix}
\left(\left(H_{02}^{21}\left(H_{02}^{11}\right)^{-1}-H_{01}^{21}\left(H_{01}^{11}\right)^{-1}\right)^{\dagger}\right)^T\\
 \left(\nul\left(H_{02}^{21}\left(H_{02}^{11}\right)^{-1}-H_{01}^{21}\left(H_{01}^{11}\right)^{-1}\right)\right)^T\end{bmatrix}^T,\label{eq:def_R}
\\
L &=  \begin{bmatrix}
\left(\left(H_{10}^{11}\right)^{-1}H_{10}^{12}-\left(H_{20}^{11}\right)^{-1}H_{20}^{12}\right)^{\dagger}\\
\nul\left(\left(H_{10}^{11}\right)^{-1}H_{10}^{12}-\left(H_{20}^{11}\right)^{-1}H_{20}^{12}\right)\\
\end{bmatrix}.\label{eq:def_L}
\end{align}
The matrices $R$ and $L$ have the following properties
\begin{align*}
&
\left(H_{02}^{21}\left(H_{02}^{11}\right)^{-1}-H_{01}^{21}\left(H_{01}^{11}\right)^{-1}\right)R = 
\begin{bmatrix} I_{N-M} \\ \mathbf{0}_{(N-M)\times (2M-N)}\end{bmatrix}^T,
\\&
L\left(\left(H_{10}^{11}\right)^{-1}H_{10}^{12}-\left(H_{20}^{11}\right)^{-1}H_{20}^{12}\right) = \begin{bmatrix}
I_{N-M}\\
\mathbf{0}_{(2M-N)\times (N-M)}
\end{bmatrix}.
\end{align*}
Note both $R$ and $L$ are full-rank $M\times M$ square matrices.
Recall that for a matrix $M$ we have $\rk(M) = \rk(R M) = \rk(M L)$
for full-rank matrices $R$ and $L$ of appropriate dimensions~\cite[eq.(0.4.6.b)]{Horn:1985:MA:5509}. So we can continue our lower bounding steps
\begin{align*}
    & \rk\left(H_{10}^{\prime}H_{12}H_{02}^{\prime}+G_{12}\right)+\rk\left(H_{20}^{\prime}H_{21}H_{01}^{\prime}+G_{21}\right)\\
   \geq  & \rk(L\left(H_{10}^{\prime}H_{12}H_{02}^{\prime}-H_{20}^{\prime}H_{21}H_{01}^{\prime}\right)R\\&+La^{12}\begin{bmatrix}
I_{N-M} & \mathbf{0}\end{bmatrix}+\begin{bmatrix}
I_{N-M}\\
\mathbf{0}
\end{bmatrix}a^{21}R\\&+
\begin{bmatrix}
I_{N-M}\\
\mathbf{0}
\end{bmatrix}a^{22}H_{02}^{21}\left(H_{02}^{11}\right)^{-1}R\\
    &+L\left(H_{20}^{11}\right)^{-1}H_{20}^{12}a^{22}\begin{bmatrix}
I_{N-M} & \mathbf{0}\end{bmatrix})\\
   = & \rk(L\left(H_{10}^{\prime}H_{12}H_{02}^{\prime}-H_{20}^{\prime}H_{21}H_{01}^{\prime}\right)R+\begin{bmatrix}
La^{12} & \mathbf{0}\end{bmatrix}\\&+
\begin{bmatrix}
a^{21}R\\
\mathbf{0}
\end{bmatrix}
+\begin{bmatrix}
a^{22}H_{02}^{21}\left(H_{02}^{11}\right)^{-1}R\\
\mathbf{0}
\end{bmatrix}
\\&+\begin{bmatrix}
L\left(H_{20}^{11}\right)^{-1}H_{20}^{12}a^{22} & \mathbf{0}\end{bmatrix})
\\=:&\rk(P).
\end{align*}
Now partition $P$ as
\begin{align*}
P  
  = & 
\begin{bmatrix}
p^{11} & p^{12}\\
p^{21} & p^{22}\\
\end{bmatrix} :
\begin{array}{l}
p^{11}\in \mathbb{C}^{\left(N-M\right)\times\left(N-M\right)},\\
p^{12}\in \mathbb{C}^{\left(N-M\right)\times\left(2M-N\right)},\\
p^{21}\in \mathbb{C}^{\left(2M-N\right)\times\left(N-M\right)},\\
p^{22}\in \mathbb{C}^{\left(2M-N\right)\times\left(2M-N\right)}.\\
\end{array}
\end{align*}
We can see that 
\begin{align*}
p^{22}  = & \nul\left(\left(H_{10}^{11}\right)^{-1}H_{10}^{12}-\left(H_{20}^{11}\right)^{-1}H_{20}^{12}\right)\\&\left(H_{10}^{\prime}H_{12}H_{02}^{\prime}-H_{20}^{\prime}H_{21}H_{01}^{\prime}\right)\\&\nul\left(H_{02}^{21}\left(H_{02}^{11}\right)^{-1}-H_{01}^{21}\left(H_{01}^{11}\right)^{-1}\right)\\
  = & \nul\left(H_{10}^{12}-H_{10}^{11}\left(H_{20}^{11}\right)^{-1}H_{20}^{12}\right)\\&\left(H_{12}-H_{10}^{11}\left(H_{20}^{11}\right)^{-1}H_{21}\left(H_{01}^{11}\right)^{-1}H_{02}^{11}\right)\\&\nul\left(H_{02}^{21}-H_{01}^{21}\left(H_{01}^{11}\right)^{-1}H_{02}^{11}\right).
\end{align*}
Because $H_{10}^{12}$ and $H_{02}^{21}$ are generic and independent of $\left(H_{12}-H_{10}^{11}\left(H_{20}^{11}\right)^{-1}H_{21}\left(H_{01}^{11}\right)^{-1}H_{02}^{11}\right)$, $p^{22}$ is full rank almost surely. Thus we have
\begin{align*}
   & \rk\left(H_{10}^{\prime}H_{12}H_{02}^{\prime}+G_{12}\right)+\rk\left(H_{20}^{\prime}H_{21}H_{01}^{\prime}+G_{21}\right)\\
  \geq & \rk\left(P\right)\\
  = & \rk\left(P/p^{22}\right)+\rk\left(p^{22}\right)\\
  \geq & \rk\left(p^{22}\right)\\
  = & 2M-N.
\end{align*}
Therefore we have 
\begin{align*}
   \dof^{\rm lin}
   \leq & 2M-\frac{2M-N}{2}
   =  M+N/2,
 \end{align*}
which implies the optimality of our scheme among all memoryless linear processing schemes at the relay.

\subsubsection{Case $N\neq L$} 
\label{sub:case_n_neq_l}
Increasing the number of antenna at the relay cannot hurt the DoF; 
we can thus always add $(L-N)^+$ antennas at the relay-receiving side and $(N-L)^+$ antennas at the relay-transmitting side,
so that the two sides have $N':=\max\{N,L\}$ antennas each. Thus for the general case we can increase the number of antennas at the relay such that it falls into one of the three cases we discussed before. As in the previous steps we show that with linear processing at the relay one necessarily has
\begin{align*}
  & \rk\left(H_{10}^{\prime}H_{12}H_{02}^{\prime}+G_{12}\right)+\rk\left(H_{20}^{\prime}H_{21}H_{01}^{\prime}+G_{21}\right)
\\\geq & 2M-\max\{N,L\},
\end{align*}
which implies
\begin{align*}
	\dof^{\rm lin} \leq M+\frac{\max\{N,L\}}{2}.
\end{align*}
We conclude that our proposed achievable scheme is optimal among all memoryless linear strategies at the relay.

\subsection{Extension to Relay with Memory} 
\label{sub:extension_to_relay_with_memory}
We consider $n$ channel extensions of the original channel. Our converse proof for the linear optimality is still valid for the extended channel. 
Thus we show the optimality of our scheme among all linear strategies at the relay, including the ones with memory.


\subsubsection{Channel Transformation} 
\label{ssub:channel_transformation_based_on_channel_coefficient_matrix}

Over $n$ channel uses the linear processing matrix $A$ at the relay would be a lower triangular matrix, as the relay can not use the received signals that it has not received yet. However, since we are interested in the upper bounds here, we can provide all received signals during $n$ channel uses to the the relay as a genie. So the encoding matrix $A$ can be an arbitrary matrix. This allow us to apply the same proof in Section \ref{sub:linear_optimality}.
The partitions of the encoding matrix $A$ and of the channel coefficient matrices remain the same as in Section \ref{ssub:channel_transformation_based_on_channel_coefficient_matrix}, except for $H_{j0}^{\prime}$	 and $H_{0j}^{\prime}$ in \eqref{eq:Hj0_prime} and \eqref{eq:H0j_prime}. 
Note that since the channel matrices are not generic anymore, $\nul\left(h_{i0}^{11}\right)$ and $\nul\left(h_{0j}^{11}\right)$ may have dimension higher than $n(M-N)$. 
Thus the definition of $H_{j0}^{\prime}$	 and $H_{0j}^{\prime}$ become
\begin{align}
H_{j0}^{\prime} & :=  \begin{bmatrix}
\left(h_{j0}^{11}\right)^{\dagger}\\
\nul_{r:n(M-L)}:\left(h_{j0}^{11}\right) \\
\end{bmatrix}:
\begin{array}{l}
\left(h_{j0}^{11}\right)^{\dagger}\in \mathbb{C}^{\min\left\{ L,M\right\} \times M},\\
\nul_{r:n(M-L)}\left(h_{j0}^{11}\right)      \in \mathbb{C}^{\left(M-L\right)^{+}\times M},\\
\end{array} 
\label{eq:Hj0_prime_n}\\
H_{0j}^{\prime} & :=  \begin{bmatrix}
\left(h_{0j}^{11}\right)^{\dagger} & \nul_{c:n(M-L)}\left(h_{0j}^{11}\right) \\
\end{bmatrix}:
\begin{array}{l}
\left(h_{0j}^{11}\right)^{\dagger}\in \mathbb{C}^{M\times \min\left\{ N,M\right\}},\\
\nul_{c:n(M-L)}\left(h_{0j}^{11}\right)      \in \mathbb{C}^{M\times\left(M-N\right)^{+}},\\
\end{array} 
\label{eq:H0j_prime_n}
\end{align}
By this definition, $H_{j0}^{\prime}$	 and $H_{0j}^{\prime}$ are still full-rank square matrices almost surely for $j\in[1:2]$.
Hence the invertible transformation in \eqref{eq:transformationwlog}
is still without loss of optimality.
The transformed channel outputs (neglecting the noise terms) can be written as \eqref{eq:transformationwlogy1} and \eqref{eq:transformationwlogy2}.
\subsubsection{Converse proof} 
\label{ssub:converser_proof}
The proof then follows exactly the same steps as in Section \ref{sub:linear_optimality}, except that all the channel matrices' dimensions are multiplied by $n$, the number of channel uses.
Note that our proof only needs several matrices to be full rank. The non-generic channel matrices do not change the results.
Specifically, the proof of cases $N=L=N$ and $N=L<M$ is  exactly the same as in Section \ref{ssub:case n=l=m} and Section \ref{ssub:case n=l<m}, respectively. 
For the case $N=L>M$, the definitions in \eqref{eq:def_R} and \eqref{eq:def_L} are changed to
\begin{align}
R &=  \begin{bmatrix}
\left(\left(H_{02}^{21}\left(H_{02}^{11}\right)^{-1}-H_{01}^{21}\left(H_{01}^{11}\right)^{-1}\right)^{\dagger}\right)^T\\
 \left(\nul_{c:n(2M-N)}\left(H_{02}^{21}\left(H_{02}^{11}\right)^{-1}-H_{01}^{21}\left(H_{01}^{11}\right)^{-1}\right)\right)^T\end{bmatrix}^T,\label{eq:def_R_n}
\\
L &=  \begin{bmatrix}
\left(\left(H_{10}^{11}\right)^{-1}H_{10}^{12}-\left(H_{20}^{11}\right)^{-1}H_{20}^{12}\right)^{\dagger}\\
\nul_{r:n(2M-N)}\left(\left(H_{10}^{11}\right)^{-1}H_{10}^{12}-\left(H_{20}^{11}\right)^{-1}H_{20}^{12}\right)\\
\end{bmatrix}.\label{eq:def_L_n}
\end{align}
Then with the new definitions the proof in Section \ref{ssub:case n=l>m} applies. 
This completes the proof of Theorem \ref{th:MainAch}.

\subsection{Discussion} 
Our results has an interesting consequence in half-duplex systems where antennas can not transmit and receive at the same time.
Assume  the relay has $M_\text{relay} := N+L$ antennas where each antenna can either send or receive.
How many antennas should we use for transmitting and how many for receiving if the goal is to maximize the sum DoF?
The answer is obtained by maximizing the achievable $\dof^{lin}$ in~\eqref{eq:achievability_sum} over $(N,L)$
subject to $N+L = M_\text{relay}$, which gives 
\begin{align*}
&
\max
\min\left\{M, N,L,\frac{\max\left\{N,L\right\}}{2}\right\}
=\min\left\{M,\frac{M_\text{relay}}{3}\right\}.
\end{align*}
For example, by optimally allocating $M_\text{relay}=2M$ antennas at the relay we obtain that
$5M/3$~DoF is achievable; this DoF is larger than $3M/2$~DoF obtained in~\cite{2ic-ir} for $L=N=M=M_\text{relay}/2$.
This implies that when the relay is restricted to linear processing, equally splitting antennas among the transmit and receive side of the relay is suboptimal.
The split ratio should be either $1/2$ or $2$.


\section{Conclusion} 
\label{sec:conclusion}

In this paper we studied the DoF of two-user multi-antenna interference channel with an instantaneous relay.
We proposed an achievable scheme that extends known results to the case where the number of transmit and receive antennas at the relay may be different.
The proposed scheme, which uses only memoryless linear operations at the relay, aims to neutralize as much interference as possible at the receivers. We identified several different regimes, depending on the relative number of antennas at the nodes.
For two such regimes (roughly speaking when either the number of receiving antennas at the relay is at least double the number of transmitting antennas, or vice-versa), we derived a information theoretic converse that shows the optimality of the proposed scheme.
For the other regimes, we showed that our proposed scheme attains the largest possible DoF among all  memoryless linear schemes at the relay (without any restriction on the processing at the transmitters).
Our result  also implies that interference alignment at the receivers cannot improve the DoF. 


So far an information theoretic converse proof is still missing for the regime $\max\{N,L\}<2\min\{M,N,L\}$. Thus the optimality of our memoryless linear scheme among all the possible strategies for this regime is open. 
When the relay has no restrictions on its processing, its output can be an arbitrary function of its input. This arbitrariness makes it difficult to compare the entropy of channel output of the relay with other terms arising in the converse.

Should our our `linear outer bound' not be tight, then linear strategies would be insufficient for this channel in terms of DoF.
For the network coding problem linear strategies are known to be insufficient in general~\cite{insufficiency_linear_NC, insufficiency_linear_IC}; these known `counter example' models 
are from complex matroids with specific linear representations and typically involve many nodes.
For the index coding problem with less than five receivers, linear strategies are optimal~\cite{YHkim.indexcodingcapacity,LOng:linearcodeICregion}.
Our channel only has 
only few nodes and thus it appears to be among those models for which linear strategies should suffice.

Since the information theoretic converse is missing for the regime $\max\{N,L\}<2\min\{M,N,L\}$, finding the optimal DoF region for our channel model is a logical next step of research.
Another future direction is to extend the analysis to more than two user pairs with one instantaneous relay;
preliminary work can be found in~\cite{Tang:MSThesis:2013}. 
The case of multiple instantaneous relays is also interesting. This would lead to a comparison among the cases of one instantaneous relay with multiple antennas (as studied in this paper) to the cases of many distributed single antenna instantaneous relays. The comparison could show the possible advantage of centralized strategy over distributed ones.
A more general setting than what considered in this paper is to have arbitrary number of antennas at each node;
finding the DoF region of this channel would show the benefit brought by the instantaneous relay to a two-user interference channel in terms of DoF.


\appendices
    
\section{The DoF Region for Gaussian two-user Interference Channel} 
\label{sec:appen_general-2ic}

Consider a two-user Gaussian IC channel with input-output relationship
\begin{align*}
Y_{1}  &=  H_{11}X_{1}+H_{12}X_{2}+Z_{1}, \\
Y_{2}  &=  H_{21}X_{1}+H_{22}X_{2}+Z_{2},   
\end{align*}
where $H_{ij}$ are of size $N_i\times M_j, \ (i,j)\in[1:2]^2$.
Note that the channel matrices are arbitrary, not generic. 
The noises $Z_{i}, i\in[0:2]$ are independent circularly-symmetric complex Gaussian vectors with zero mean and covariance matrix equal to the identity. The inputs are subject to power constraint
\begin{align*}
	X_i\in \mathbb{C}^{M_i\times 1}: \text{Tr}[\Sigma_i]\leq \mathsf{P},
\end{align*}
where $\Sigma_i$ is the covariance matrix of input $X_i, \ i\in[1:2]^2$.
Achievable rates, capacity region and DoF region are defined in the usual way~\cite{NIT}.

Our main result in this section is:
\begin{thm}
\label{thm:general_dof_region}
The DoF region for the two-user Gaussian IC is
\begin{subequations}
\begin{align}
	d_1 &\leq \rk(H_{11}),\label{eq:general_dof_region d1}
	\\
	d_2 &\leq \rk(H_{22}),\label{eq:general_dof_region d2}
	\\
	d_1+d_2 &\leq \rk\left( \begin{bmatrix}
		H_{11} & H_{12}	\end{bmatrix}
		\right)
		+\rk\left(
		\begin{bmatrix}
			H_{12} \\ H_{22}
		\end{bmatrix}
		\right)
		-
		\rk\left(H_{12}\right),
		\label{eq:general_dof_region d1+d2 kra1}
		\\
	d_1+d_2 &\leq \rk\left( \begin{bmatrix}
		H_{2} & H_{21}	\end{bmatrix}
		\right)
		+\rk\left(
		\begin{bmatrix}
			H_{11} \\ H_{21}
		\end{bmatrix}
		\right)
		-
		\rk\left(H_{21}\right),
		\label{eq:general_dof_region d1+d2 kra2}
		\\
	d_1+d_2 &\leq \rk\left( \begin{bmatrix}
		H_{11} & H_{12}\\ H_{21} & 0	\end{bmatrix}
		\right)
		+\rk\left(
		\begin{bmatrix}
			0 & H_{12} \\ H_{21} & H_{22}
		\end{bmatrix}
		\right)
		-
		\rk\left(H_{12}\right)-\rk(H_{21}),
		\label{eq:general_dof_region d1+d2 etw}
		\\
	2d_1+d_2 &\leq  \rk\left( \begin{bmatrix}
		H_{11} & H_{12} \end{bmatrix}
		\right)
		+\rk\left(
		\begin{bmatrix}
			H_{11} \\ H_{21} 
		\end{bmatrix}
		\right)
		+\rk\left(
		\begin{bmatrix}
			0 & H_{12} \\ H_{21} & H_{22}
		\end{bmatrix}
		\right)
		-
		\rk\left(H_{12}\right)-\rk(H_{21}),
		\label{eq:general_dof_region 2*d1+d2}
		\\
	d_1+2d_2 &\leq  \rk\left( \begin{bmatrix}
		H_{22} & H_{21} \end{bmatrix}
		\right)
		+\rk\left(
		\begin{bmatrix}
			H_{22} \\ H_{12} 
		\end{bmatrix}
		\right)
		+\rk\left(
		\begin{bmatrix}
			0 & H_{21} \\ H_{12} & H_{11}
		\end{bmatrix}
		\right)
		-
		\rk\left(H_{12}\right)-\rk(H_{21}).
		\label{eq:general_dof_region d1+2*d2}
\end{align}	
\label{eq:general_dof_region}
\end{subequations}
\end{thm}
\begin{IEEEproof}
For two-user Gaussian IC the following upper bound is optimal to within a constant gap~\cite{IC_constant_gap}
\begin{align*}
	R_1 &\leq I(X_1;Y_1|X_2,Q) &\text{cut-set bound},\\
	R_2 &\leq I(X_2;Y_2|X_1,Q) &\text{cut-set bound},\\
	R_1+R_2 &\leq I(X_1,X_2;Y_1,Q)+I(X_2;Y_2|X_1,Y_1,Q)   &\text{from~\cite{D.Tuninetti:gaussian_outer_bound}},\\
	R_1+R_2 &\leq I(X_1,X_2;Y_2,Q)+I(X_1;Y_1|X_2,Y_2,Q)   &\text{from~\cite{D.Tuninetti:gaussian_outer_bound}},\\
	R_1+R_2 &\leq h(Y_1|U_1,Q)+h(Y_2|U_2,Q)-h(Z_1)-h(Z_2) &\text{from~\cite{IC_constant_gap}},\\
	2R_1+R_2 &\leq h(Y_1|U_1,X_1,Q)+h(Y_1|Q)+h(Y_2|U_2,Q)-2h(Z_1)-h(Z_2) &\text{from~\cite{IC_constant_gap}},\\
	R_1+2R_2 &\leq h(Y_2|U_2,X_2,Q)+h(Y_2|Q)+h(Y_1|U_1,Q)-h(Z_1)-2h(Z_2) &\text{from~\cite{IC_constant_gap}},
\end{align*} 
for some $P_{X_1X_2Q}=P_{Q}P_{X_1|Q}P_{X_2|Q}$, and where 
\begin{align*}
U_1 &:=H_{21}X_1+Z_2^{\prime} \\
U_2 &:=H_{12}X_2+Z_1^{\prime} \\
&(Z_1^{\prime},Z_2^{\prime}) \ \text{independent copies of} \ (Z_1,Z_2).
\end{align*}
Thus we can derive the DoF region based on it.

Specifically, from the well known single-user MIMO bound we have 
\begin{align*}
R_1 &\leq I(X_1;Y_1|X_2,Q)\\
    &\leq I(H_{11}X_1+Z_1;X_1)\\
	&\leq \max_{0\preceq \Sigma_1 :  \text{Tr}[\Sigma_1]\leq \mathsf{P}}\log |H_{11}\Sigma_1 H_{11}^{H}+I|\\
	&\leq \max_{0\preceq \Sigma_1 \preceq I \mathsf{P}} \log |H_{11}\Sigma_1 H_{11}^{H}+I|\\
	&= \log |H_{11} H_{11}^{H} \ \mathsf{P}+I|\\
	&= \rk(H_{11}) \log(1+\mathsf{P})+O(1) \ \text{as $\mathsf{P}\to\infty$}.
\end{align*}
Thus we have $d_1 \leq \rk(H_{11})$ in~\eqref{eq:general_dof_region d1}.
Similarly, we have  $d_2 \leq \rk(H_{22})$  in~\eqref{eq:general_dof_region d2}.

For the sum-rate we have
\begin{align*}
R_1+R_2 
	&\leq I(X_1,X_2;Y_1)+I(X_2;Y_2|X_1,Y_1)\\
	&= I(X_1,X_2;H_{11}X_1+H_{12}X_2+Z_1)\\&  +I(X_2;H_{22}X_2+Z_2|H_{12}X_2+Z_1)\\
	&= h(H_{11}X_1+H_{12}X_2+Z_1)-h(Z_1) \\&  +h(H_{22}X_2+Z_2,H_{12}X_2+Z_1)-h(H_{12}X_2+Z_1)-h(Z_2)
	\\
	&\leq \max_{0\preceq \Sigma_i :  \text{Tr}[\Sigma_i]\leq \mathsf{P}, i\in[1:2]}
	h\left(\begin{bmatrix}
		H_{11} & H_{12}
	\end{bmatrix}
	\begin{bmatrix}
		X_1 \\ X_2
	\end{bmatrix}+Z_1\right)-h(Z_1)
	\\&\qquad
	+h\left(\begin{bmatrix}
		H_{22} \\ H_{12}	
	\end{bmatrix}X_2+\begin{bmatrix}
			Z_2 \\ Z_1
	\end{bmatrix}\right)-h(H_{12}X_2+Z_1)-h(Z_2)
	\\
	&\leq \left(\rk\left(\begin{bmatrix}
		H_{11} & H_{12}
	\end{bmatrix}\right)+\rk\left(\begin{bmatrix}
		H_{12}\\ H_{22}
	\end{bmatrix}\right)-\rk(H_{12})\right) \log(1+\mathsf{P})+O(1) \ \text{as $\mathsf{P}\to\infty$},
\end{align*}
where the last inequality follows from Gaussian maximizing the differential entropy and extremal inequality in \cite{extremal_inequ}.
This shows the bound in~\eqref{eq:general_dof_region d1+d2 kra1}, and by swapping the role of the users the one in~\eqref{eq:general_dof_region d1+d2 kra2}.

For the bound in~\eqref{eq:general_dof_region d1+d2 etw}, we have
\begin{align*}
	R_1+R_2 &\leq h(H_{11}X_1+H_{12}X_2+Z_1|H_{21}X_1+Z_2^{\prime})\\&\quad+h(H_{21}X_1+H_{22}X_2+Z_2|H_{12}X_2+Z_1^{\prime})-h(Z_1)-h(Z_2)\\
	&= h\left(
	\begin{bmatrix}
		H_{11} & H_{12}\\ H_{21} & 0
	\end{bmatrix}
	\begin{bmatrix}
		X_1 \\ X_2
	\end{bmatrix}
	+
	\begin{bmatrix}
		Z_1 \\ Z_2^{\prime}
	\end{bmatrix}
	\right)
	-h(H_{21}X_1+Z_2^{\prime})
	\\&\qquad+
	h\left(
	\begin{bmatrix}
		H_{21} & H_{22} \\ 0 & H_{12}
	\end{bmatrix}
	\begin{bmatrix}
		X_1 \\ X_2
	\end{bmatrix}+
	\begin{bmatrix}
		Z_2 \\ Z_1^{\prime}
	\end{bmatrix}
	\right)-h(H_{12}X_2+Z_1^{\prime})-h(Z_1)-h(Z_2)\\
	&= h\left(
	\begin{bmatrix}
		H_{11} & H_{12}\\ H_{21} & 0
	\end{bmatrix}
	\begin{bmatrix}
		X_1 \\ X_2
	\end{bmatrix}
	+
	\begin{bmatrix}
		Z_1 \\ Z_2^{\prime}
	\end{bmatrix}
	\right)-
	h\left(
	\begin{bmatrix}
		H_{21} & 0\\ 0 & H_{12}
	\end{bmatrix}
	\begin{bmatrix}
		X_1 \\ X_2
	\end{bmatrix}
	+
	\begin{bmatrix}
		Z_2^{\prime} \\ Z_1^{\prime}
	\end{bmatrix}
	\right)
	\\&\qquad+
	h\left(
	\begin{bmatrix}
		H_{21} & H_{22} \\ 0 & H_{12}
	\end{bmatrix}
	\begin{bmatrix}
		X_1 \\ X_2
	\end{bmatrix}+
	\begin{bmatrix}
		Z_2 \\ Z_1^{\prime}
	\end{bmatrix}
	\right)-h(Z_1)-h(Z_2)\\
	&\leq \left(
	\rk \left(
	\begin{bmatrix}
		H_{11} & H_{12} \\ H_{21} & 0 
	\end{bmatrix}
	\right)
	-\rk(H_{21})-\rk(H_{12})+\rk\left(
	\begin{bmatrix}
		H_{21} & H_{22} \\ 0 & H_{12}
	\end{bmatrix}
	\right)
	\right)
	 \log(1+\mathsf{P})\\&\qquad+O(1) \ \text{as $\mathsf{P}\to\infty$},
\end{align*}
where the last inequality follows from follows from Gaussian maximizing the differential entropy and extremal inequality in \cite{extremal_inequ}.
Similarly, one can show the bound in~\eqref{eq:general_dof_region 2*d1+d2} and~\eqref{eq:general_dof_region d1+2*d2}.

This concludes the proof of Theorem \ref{thm:general_dof_region}.
\end{IEEEproof}

\bibliographystyle{IEEEtranS}
\bibliography{refs}

\end{document}